\documentclass[preprint,3p,twocolumn,number]{elsarticle}
\journal{Chaos, Solitons \& Fractals}

\usepackage{float,amsmath,hyperref}

\begin{document}
	
\begin{frontmatter}
	
\title{Topological phase transition in the periodically forced Kuramoto Model}

\author[add1]{E.~A.~P.~Wright\corref{cor1}}
\cortext[cor1]{Please address correspondence to E.~A.~P.~Wright}
\ead{wrighteap@ua.pt}

\author[add1]{S.~Yoon}
\author[add1]{J. F. F. Mendes}
\author[add1,add2]{A. V. Goltsev}

\address[add1]{Department of Physics $\&$ I3N, University of Aveiro, 3810-193 Aveiro, Portugal}
\address[add2]{A.F. Ioffe Physico-Technical Institute, 194021 St. Petersburg, Russia}

\begin{abstract}
A complete bifurcation analysis of explicit dynamical equations for the periodically forced Kuramoto model was performed in [L. M. Childs and S. H. Strogatz. Chaos \textbf{18}, 043128 (2008)], identifying all bifurcations within the model. We show that the phase diagram predicted by this analysis is incomplete. Our numerical analysis of the equations reveals that the model can also undergo an abrupt phase transition from oscillations to wobbly rotations of the order parameter under increasing field frequency or decreasing field strength. This transition was not revealed by bifurcation analysis because it is not caused by a bifurcation, and can neither be classified as first {nor} second order since it does not display critical phenomena characteristic of either transition. We discuss the topological origin of this transition and show that it is determined by a singular point in the order-parameter space.
\end{abstract}

%\begin{highlights}
%\item We reveal a topological phase transition in the periodically forced Kuramoto model
%\item The transition separates states with oscillating and rotating collective dynamics
%\item The critical point is accompanied by an abrupt drop in the average cycling frequency
%\item Oscillating and rotating states have distinct winding numbers
%\item Winding numbers are defined relative to a singular point in the order-parameter space
%\end{highlights}


\end{frontmatter}

%\maketitle

\section{\label{sec:Introdution} Introduction}

The Kuramoto model describes synchronization between coupled phase oscillators, and has been studied extensively since it was first introduced \cite{kuramoto1975international}. This representative model has attracted much attention for its wide range of applications in physics, neuroscience, biology, and the social sciences as well as the variety of collective dynamical phenomena it demonstrates \cite{Str2000Kuramoto,acebron2005kuramoto,Are2008Synchronization,Rod2016Kuramoto}, including exotic collective states such as chimera states \cite{abrams2004chimera} and Bellerophon states \cite{qiu2016synchronization}, well known first and second order phase transitions \cite{Str2000Kuramoto,acebron2005kuramoto,Are2008Synchronization,Rod2016Kuramoto}, and an unusual hybrid phase transition \cite{coutinho2013kuramoto}. Here, from among the different lines of enquiry, we consider the problem of entrainment by an external periodic field. The phase diagram for a system of uniformly coupled Kuramoto oscillators in an external periodic field was first reported in \cite{sakaguchi1988cooperative}.
%, revealing how the spread of oscillator frequencies and the field strength determine if synchronized states are entrained to the field frequency or disrupted. Later work identified conditions under which the macroscopic dynamics of the model are described by a system of two coupled differential equations \cite{antonsen2008external,ott2008low}, enabling a detailed phase diagram of the model based on a complete bifurcation analysis \cite{childs2008stability}.
{More recently, Childs and Strogatz performed a} complete bifurcation analysis of explicit dynamical equations {for the} periodically forced Kuramoto model{, revealing the} exact locations of all bifurcations \cite{childs2008stability}.

In this work, we report a new phase transition in the periodically forced Kuramoto model, based on numerical and analytical analysis of the explicit dynamical equations derived in \cite{childs2008stability}. We show that this transition cannot be revealed by bifurcation analysis because it is not caused by a bifurcation. Moreover, it {cannot} be classified as a first or second order phase transition since it does not display critical phenomena characteristic of either transition. {Our analysis reveals the topological character of this transition, and the crucial role of a singular point in the order-parameter space.} Topological phase transitions {have been} observed and studied in quantum systems such as fractional quantum Hall liquids and  topological isolators \cite{wen1995topological,hasan2010colloquium,goldman2016topological}.  Our work shows that a topological transition can also occur in a classical system such as the forced Kuramoto model.

In the next section, we provide an overview of the forced Kuramoto model, the order-parameter space, and a brief summary of the {physical behavior displayed by the model}. In Section~\ref{sec:three}, we present an updated phase diagram and our analysis of the transition. Finally, we discuss our findings in Section~\ref{sec:four}.

\section{\label{sec:two} The Forced Kuramoto Model}

In the Kuramoto model with a periodic field of strength \(F\) and angular frequency \(\sigma\), the phase \(\theta_i\) of the \(i^\text{th}\) phase oscillator is given by
\begin{equation}
  \frac{d \theta_i}{d t} = \omega_i + \frac{K}{N}\sum_{j=1}^{N}\sin(\theta_j-\theta_i) + F\sin(\sigma t - \theta_i),
  \label{eq:km}
\end{equation}
where \(\omega_i\) is the oscillator's natural frequency, \(K\) is the uniform coupling strength, and \(N\) the number of oscillators in the system. The macroscopic state of the system is conveniently described by the complex order parameter
\begin{equation}
  z = \rho e^{i\psi} = \frac{1}{N} \sum_{j=1}^{N} e^{i\theta_j},
  \label{eq:order_parameter}
\end{equation}
where the parameter $\rho$ characterizes the degree of synchronization between the oscillators ($0 \leq \rho \leq 1$) and the phase $\psi$ shows the direction of alignment. Applying the Ott-Antonsen ansatz \cite{ott2008low} to a system of oscillators with natural frequencies drawn from a Lorentzian distribution of spread \(\Delta=1\), in a frame rotating at the field frequency $\sigma$, the macroscopic dynamics of the system are reduced to
\begin{align}
\frac{d \rho}{d t} &= -\rho + \frac{K}{2} \rho (1{-}\rho^2) {+} \frac{F}{2} (1{-}\rho^2) \cos\psi,
\label{eq:rho} \\
\frac{d \psi}{d t} &= -\Omega - \frac{F}{2} \frac{(1+\rho^2)}{\rho} \sin\psi,
\label{eq:psi}
\end{align}
where \(\Omega=\sigma - \omega_0\) is the detuning parameter, and \(\omega_0\) is the average natural frequency of oscillators. Note that {the resulting} equations are symmetric with respect to the replacement $\Omega \rightarrow -\Omega$ and $\psi \rightarrow - \psi$. For simplicity, we will assume \(\omega_0=0\), so that \(\Omega=\sigma\). {Equivalently}, one {may} consider a frame rotating with frequency \(\omega_0\){, where} the field frequency is $\sigma-\omega_0$.

Eqs.~(\ref{eq:rho}) and (\ref{eq:psi}) describe the motion of the system in a two-dimensional order-parameter space
%or manifold of internal states
$(\text{Re} (z), \text{Im} (z))=(\rho \cos \psi, \rho \sin \psi)$
%, where each internal state is defined by polar coordinates $\rho$ and $\psi$,
{where} positive rotation of $\psi$ is counterclockwise. It is important to note that the state $z=0$ is a singular point in this space because $\psi$ becomes undefined when $\rho=0$. Since any trajectory through this order-parameter space requires the function $\psi(t)$ to be analytic at every time $t$, e.g. the first derivative defines angular velocity, any trajectory through the singular point $z=0$ is forbidden.

The complete bifurcation analysis of Eqs. $(\ref{eq:rho})$ and $(\ref{eq:psi})$ was performed by Childs and Strogatz in \cite{childs2008stability}  where one can find the explicit stability diagram, all bifurcations curves, and phase portraits of dynamical states of the forced Kuramoto model at different model parameters $F, \Omega$ and $K$. In short, the model demonstrates the following physical behavior.
First, in the absence of the field, when the coupling $K$ between phase oscillators is larger than the critical coupling $K_c =2$, interaction results in a synchronous rotation of phase oscillators with the group velocity $\omega_0$ and order parameter $\rho=\sqrt{(K-2)/K}$. %The rotational symmetry is spontaneously broken and the phase $\psi$, along which oscillators' phases $\theta_i$ are aligned is an arbitrary.
At $K=K_c$, the rotational symmetry is spontaneously broken and {the phases \(\theta_i\) of a finite fraction of oscillators become aligned along an arbitrary direction $\psi$.} %Second, in applied field $F$, in a broad region of the model parameters, when $K >2$ and $|\Omega|$ is sufficiently smaller than $F$, the system is in the entrained state (see phase I in Fig. \ref{fig:phase_diagram}). 
Second, {upon applying a field with strength} $F$ {at detuning $|\Omega|$ sufficiently smaller than $F$, the synchronized group becomes entrained by the field, for a broad range of model parameters (see phase I in Fig. \ref{fig:phase_diagram}).} In the original non-rotating frame, the entrained state is phase- and frequency-locked to the field and a finite fraction of phase oscillators rotates synchronously with the field frequency $\sigma$.
%In the rotating frame, the entrained state is a stationary solution of Eqs. $(\ref{eq:rho})$ and $(\ref{eq:psi})$ and corresponds to a stable fixed point.
If $\Omega > 0$, the order parameter $z$ lags behind the field, i.e., the phase $\psi$ is negative and the order parameter $z$ lies in the lower half-plane of the order-parameter space.  If $\Omega < 0$, then the order parameter $z$ is ahead of the field, the phase $\psi$ is positive and $z$ lies in the upper half-plane. {The latter symmetry} follows from the symmetry of Eqs. $(\ref{eq:rho})$ and $(\ref{eq:psi})$ with respect to the replacement $\Omega \rightarrow -\Omega$ and $\psi \rightarrow - \psi$. {Third, increasing $\Omega$ at fixed \(F\) disrupts the entrained state and leads to periodic dynamics in the rotating frame, as the system undergoes a SNIPER, saddle-node, or Hopf bifurcation \cite{childs2008stability}. As discussed below, we found that when entrainment is disrupted, coupled oscillators can either oscillate synchronously with respect to the field direction (phase II in Fig.~\ref{fig:phase_diagram}) or drift at a frequency other than the field frequency (phase III in Fig.~\ref{fig:phase_diagram}).}

\section{SPOR transition \label{sec:three}}

\begin{figure}[H]
	\includegraphics[width=\linewidth]{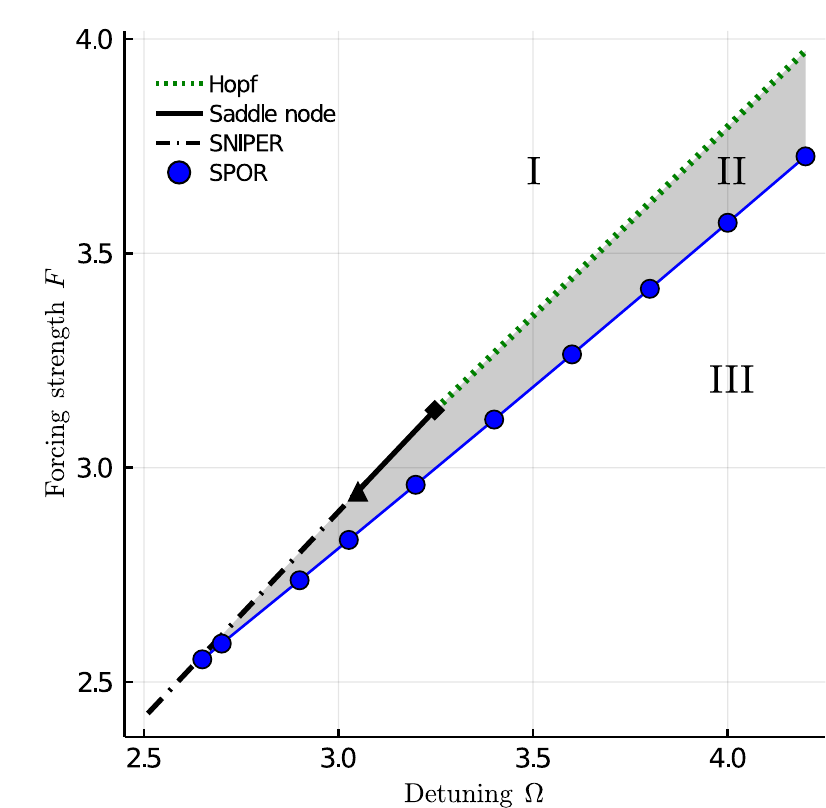}
	\caption{\label{fig:phase_diagram} Partial phase diagram for the Kuramoto model in a homogenous field with strength \(F\) and detuning \(\Omega\) at \(K=5\). Each phase is characterized by distinct dynamics of \(\psi(t)\). Phase \(\mathrm{I}\) corresponds to the entrained phase. Phase \(\mathrm{II}\) is the oscillating phase. Phase \(\mathrm{III}\), is the rotating phase. Full circles (blue) define a boundary (the SPOR transition) separating phases \(\mathrm{II}\) (shadowed region) and \(\mathrm{III}\). The upper boundary of phase \(\mathrm{I}\) was calculated following \cite{childs2008stability}: the dash-dotted line (black) indicates a SNIPER bifurcation, which becomes a Saddle-Node bifurcation (full black line) at the triangular marker, which in turn becomes a Hopf bifurcation (dotted green line) at the square marker. A detailed phase diagram in the region I is presented in \cite{childs2008stability}.}
\end{figure}

Our numerical analysis of Eqs. $(\ref{eq:rho})$ and $(\ref{eq:psi})$ revealed that the phase diagram in \cite{childs2008stability} is incomplete since it misses a phase transition from the phase with oscillations (phase II in Fig. \ref{fig:phase_diagram}) into a phase where the phase oscillators demonstrate a  wobbling rotation around the singular point $z=0$ with an angular frequency that differs from the field frequency (phase III in Fig. \ref{fig:phase_diagram}). In order to understand the origin and properties of this transition, which we have tentatively named a Singular Point Oscillation-to-Rotation (SPOR) transition, we performed a detailed numerical analysis of Eqs. $(\ref{eq:rho})$ and $(\ref{eq:psi})$.

\begin{figure}[H]
	\includegraphics[width=\linewidth]{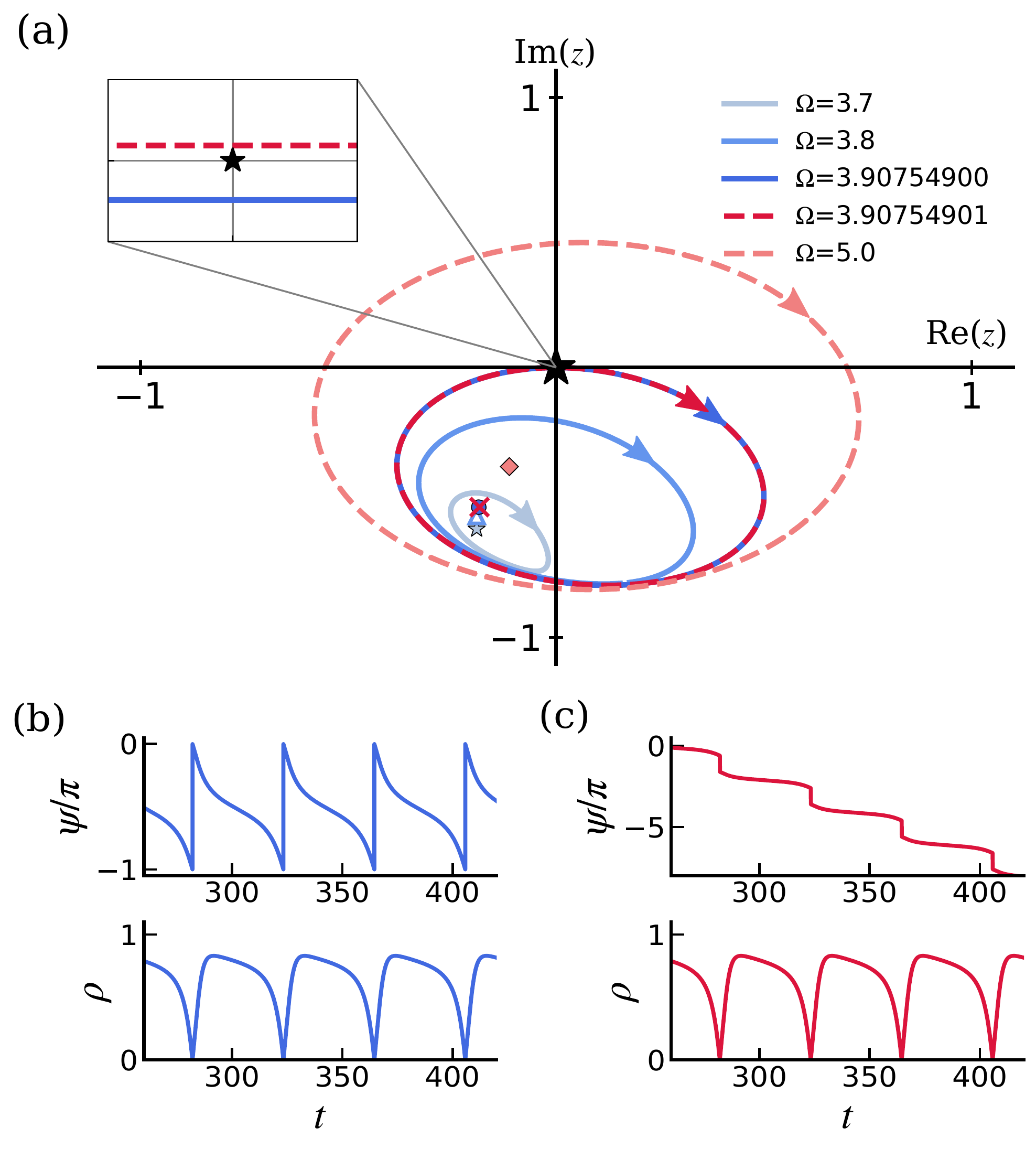}
	\caption{(a) Orbits of the order parameter \(z\) at different $\Omega$ and parameters $F=3.5$ and $K=5$. Dynamics of $\psi$ and $\rho$ in (b) the oscillating phase at $\Omega = 3.90754900$ below the SPOR transition, and (c) the rotating phase at $\Omega = 3.90754901$ above the SPOR transition. Our estimation of the critical detuning parameter is $3.90754900 <\Omega_C < 3.90754901$. The origin of the complex plane (singular point) is marked with a star. The remaining markers indicate the unstable fixed point (spiral) and are mapped to each orbit by color. The inset in (a) shows the region of the order parameter space near the singular point, and a clear change in the position of the orbits relative to the singular point for a small change (\(10^{-8}\)) in $\Omega$. Arrowheads in (a) indicate the direction of motion along the orbits. \label{fig:phase_portrait}}
\end{figure}

First, we studied the oscillations in phase II, by fixing the field strength $F$ and increasing $\Omega$. The non-uniform dynamics of \(z\) {in phase II} are typified in Fig.~\ref{fig:phase_portrait}(b), for oscillations born from a Hopf bifurcation. In this phase, the order parameter \(z\) describes a limit cycle in the lower half plane of the order-parameter space. The amplitude of the oscillations in $\psi$ is bound between \(\psi_{max}\) and \(\psi_{min}\), so that  \(\psi_{max} - \psi_{min}<\pi\). As the detuning \(\Omega\) increases, the amplitude \(\psi_{max} - \psi_{min}\) tends to \(\pi\) at a critical detuning \(\Omega_C\), which determines the boundary between phases II and III. Fig.~\ref{fig:phase_portrait}(b) shows that over one period of oscillation, the order parameter slowly falls behind the field ($\psi$ moves slowly from 0 to $-\pi$) and then quickly catches up ($\psi$ quickly moves from $-\pi$  to 0). The sharp increase of $\psi$ from a value a little bit above $-\pi$ to a value a little bit below 0 corresponds to fast motion of $z$ along the upper part of the limit cycle, due to high angular velocity $d \psi / dt$ caused by the $1/\rho$ singularity in Eq.~$(\ref{eq:psi})$. Finally, we note that in the rotating frame the angular velocity $d \psi / dt$  averaged over the period of oscillations is zero. Thus,  \(z\) is on average frequency-locked to the field. At \(\Omega> \Omega_C\), {we enter} phase III,  {and} the order parameter \(z\) starts rotating clockwise {in the rotating frame}, around the singular point $z=0$, as shown in Figs.~\ref{fig:phase_portrait}(a) and \ref{fig:phase_portrait}(c). Note that the parameter $\rho$ has the same behavior above and below $\Omega_C$ (see Figs.~\ref{fig:phase_portrait}(b) and \ref{fig:phase_portrait}(c)).

\begin{figure}[H]
	\includegraphics[width=\linewidth]{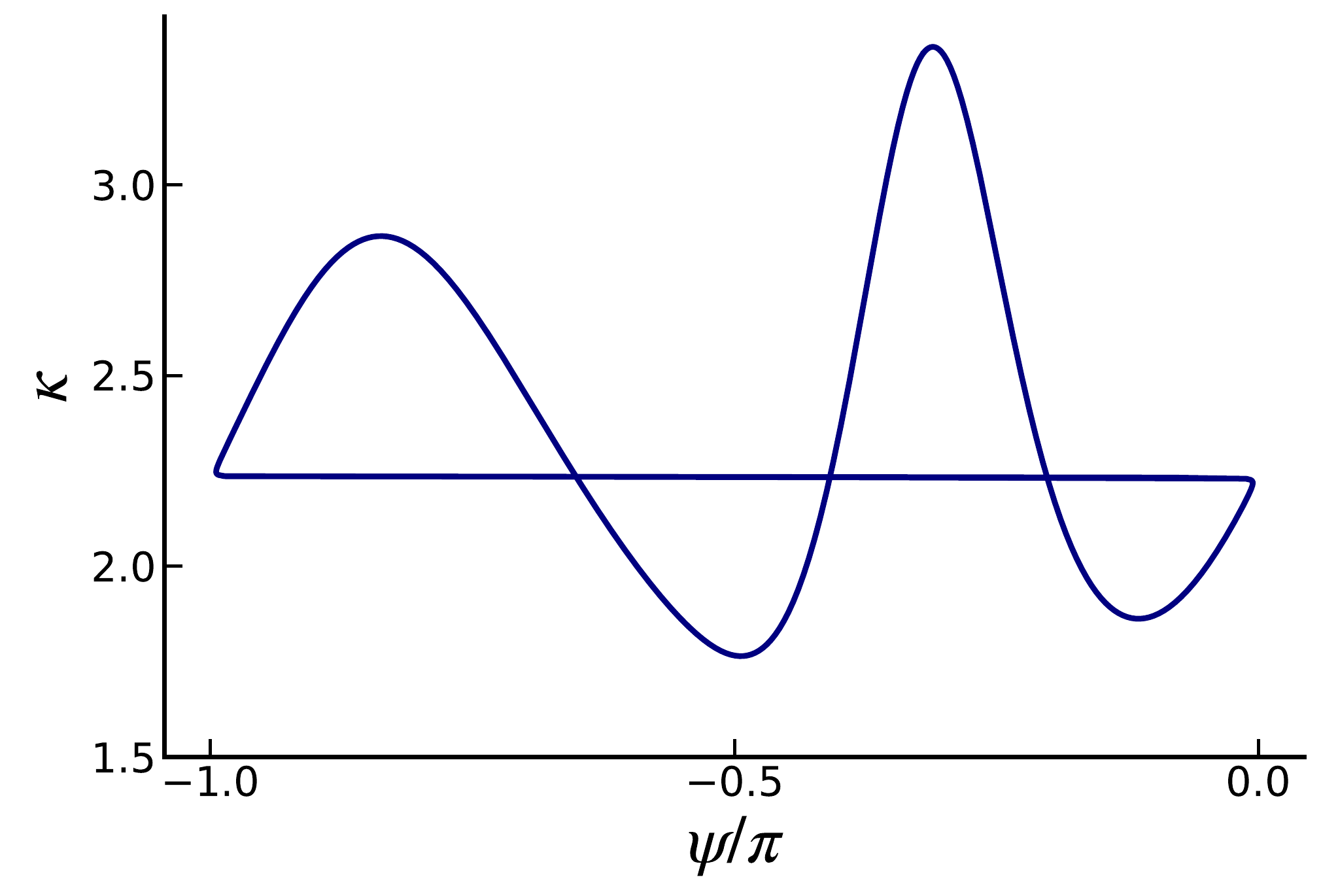}
	\caption{\label{fig:curvature} Curvature $\kappa$ versus $\psi$ of the limit  cycle of oscillations near the critical point of the transition to rotation. Parameters: $K=5$, $F=3.5$, $\Omega=3.907549$.}
\end{figure}

In order to understand the origin of the SPOR transition, we analyzed the behavior of the bifurcation point and limit cycle. In phases II (oscillating phase) and III (rotating phase) of the phase diagram there is only one bifurcation point, an unstable spiral. In the oscillating phase, the limit cycle lies in the lower half-plane as shown in Fig.~\ref{fig:phase_portrait}(a). Increasing $\Omega$ moves the system away from the critical boundary with region I, {expanding the limit cycle around the bifurcation point, and moving the upper part of the limit cycle towards the singular point $z=0$. For $\Omega < \Omega_C$, the limit cycle remains in the lower half-plane, never crossing the singular point. Increasing $\Omega$ from below to above $\Omega_C$ causes the bifurcation point to move continuously towards the singular point $z=0$, as depicted in Fig.~\ref{fig:phase_portrait}(a).} Next, we numerically analyzed the shape and curvature of the limit cycles and found no peculiarities even at $\Omega$ very close \((10^{-8}\)) to $\Omega_C$.
We used the well known equation for the curvature $\kappa$ of a curve, given by a function $\rho=\rho(\psi)$  in polar coordinates $(\rho, \psi)$:
\begin{equation}
\kappa(\psi)=\frac{|\rho^2 +2(\rho')^2 - \rho\rho''|}{[\rho^2 + (\rho')^2]^{3/2}},
\label{eq:curv}
\end{equation}
where $\rho'\equiv d\rho /d\psi$. We found the function $\rho(\psi)$, which defines the limit cycle, and its derivatives by use of the numerical solution $\rho(t)$ and $\psi(t)$. In the tested range of \(\Omega\) below $\Omega_C$, the limit cycles presented no peculiarity in shape, such as flattening (which corresponds to $\kappa \rightarrow 0$). The curvature is finite and nonzero at all points on the limit cycle, as shown in Fig. \ref{fig:curvature} for the limit cycle of oscillations at $\Omega$ very close to $\Omega_C$. 
\noindent We also extended our numerical analysis to limit cycles above the critical point $\Omega_C$. At \(\Omega > \Omega_C\), the singular point is encircled by the limit cycle, as shown in Fig.~\ref{fig:phase_portrait}(a). The shape of the limit cycle, its curvature, and the tangential velocity are similar to those below $\Omega_C$. Although we found the angular velocity diverges in the vicinity of the singular point $z=0$ as \(\Omega\) tends to $\Omega_C$, namely that $\max(d\psi/dt) \rightarrow + \infty$ from below and $\min(d\psi/dt) \rightarrow - \infty$ from above, the tangential velocity $(dx/dt, dy/dt)$ is finite at all points on the limit cycle, where $x=\text{Re} (z)$ and $y=\text{Im} (z)$. Finally, we analyzed the relaxation time above and below  $\Omega_C$  and found that the relaxation time is finite and does not demonstrate critical behavior such as critical slowing down. Likewise, we found no evidence of hysteresis depending on initial conditions. Thus, we found no evidence suggesting the observed transition is second- or first-order, as would otherwise be suggested by the presence of the above-mentioned phenomena.

\begin{figure}[H]
	\includegraphics[width=\linewidth]{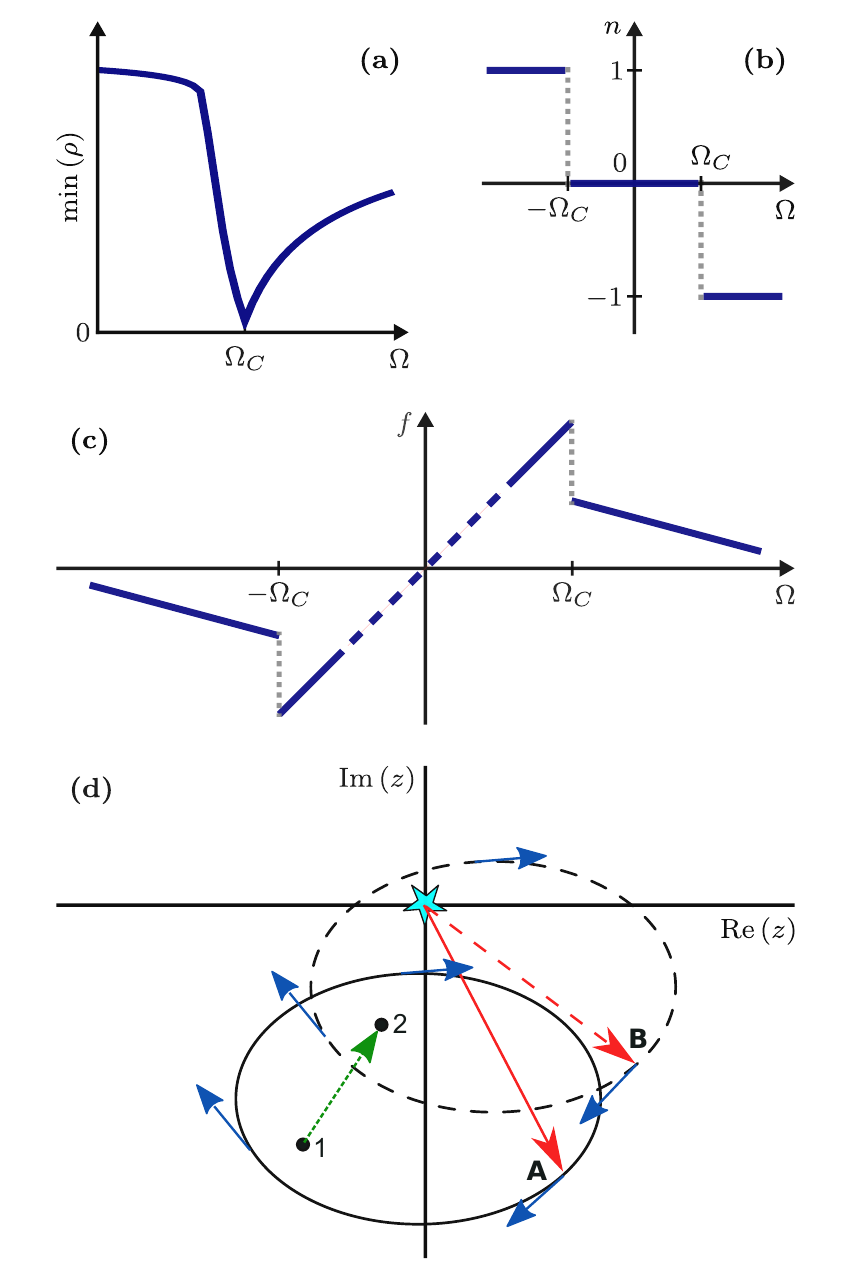}
	\caption{(a) Minimum value of the amplitude \(\rho\left(t\right)\) of the complex order parameter \(z=\rho\exp(i\psi)\) as a function of the detuning \(\Omega\). (b) Winding number $n$, Eq. \ref{eq:winding_number}, of the orbit described by \(z\) in a frame rotating at the field frequency, for both positive and negative \(\Omega\). (c) Average cycling frequency of rotation \(f\) in the original {non-rotating} frame, for both positive and negative \(\Omega\). The dashed line indicates non-oscillating phases, and \(\Omega_C\) is the critical detuning. (d) Schematic representation of two orbits for detuning \(\Omega\) slightly greater (dashed line) and slightly smaller (solid line) than \(\Omega_C\). These orbits differ simply by the translation represented by the green dashed vector connecting unstable spiral points 1 and 2, and the translation is exaggerated for illustrative purposes. The tangential velocity at equivalent points under the translation is the same for both orbits, as shown at points A and B. The red vectors from the origin of the complex plane (marked with a star) to points A and B describe orbits with winding numbers 0 (solid vector) and \(-1\) (dashed vector). \label{fig:figure_4}}
\end{figure}

{In the original {non-rotating} frame, the phase of the order parameter \(z\) equals $\widetilde{\psi}=\sigma t + \psi$. Therefore, the average cycling frequency $f$ of \(z\) over one period \(T\) of oscillations or wobbling rotations is
\begin{equation}
     f = \frac{1}{2\pi T}\int_{t}^{t+T} dt' \frac{d \widetilde{\psi}}{dt'} = \frac{\sigma}{2\pi} + \frac{1}{2 \pi T} \left[ \psi(t+T)- \psi(t) \right],
     \label{eq:avg_frq}
\end{equation}
where $t$ is an arbitrary point in time, and the period $T$ is a function of $\Omega$, $F$, and $K$. From Eq.~(\ref{eq:avg_frq}), it follows that $f= \sigma/(2\pi )$ in the oscillating phase, where $\psi(t+T)= \psi(t)$, showing the system is on average frequency-locked to the field. In the rotating phase, the average cycling frequency is $f= \sigma/(2\pi )- 1/T$ since $\psi(t+T)= \psi(t)-2\pi$ for clockwise rotation, and the frequency $f$ tends to $\omega_0$ for increasing \(\Omega\) since a high frequency field does not impact coupled phase oscillators. Thus, the SPOR transition from oscillations to wobbling rotations appears as an abrupt drop in the average cycling frequency  $f$ at $\Omega_C$, and the cycling frequency  $f$ decreases for $\Omega > \Omega_C$, as shown in Fig. \ref{fig:figure_4} (c). The drop equals the inverse of the oscillation period $T$ at the critical point.} Our numerical results allow us to conclude that the only important difference between limit cycles below and above  $\Omega_C$ is in the relative position of the singular point $z=0$. The singular point lies outside the limit cycles at \(\Omega < \Omega_C\) and is encircled by the limit cycles at \(\Omega > \Omega_C\), as shown in Fig.~\ref{fig:phase_portrait}(a). As a result, these two types of limit cycles, each associated with distinct dynamical states, acquire different topological properties. Based on the topological theory of ordered media \cite{mermin1979topological}, we can characterize limit  cycles by a winding number $n$. The field $\textbf{s}(\textbf{r})=(\rho \cos \psi, \rho \sin \psi)$ is known on all points $\textbf{r}$ on a limit  {cycle} [see, for example, vectors $\textbf{A}$ and $\textbf{B}$ in Fig. \ref{fig:figure_4}(d)]. We can measure the total angle $\psi$ when the vector $\textbf{s}(\textbf{r})$
turns as $\textbf{r}$ traverses the complete limit cycle (counterclockwise increments in angle are positive and clockwise increments are negative). Since $\textbf{s}(\textbf{r})$  is continuous on the {cycle}, this angle must be an integral multiple of $2\pi$. 

\noindent In addition, since we are mapping the temporal behavior of the order  parameter $z(t)$ to the order-parameter space $(\rho, \psi)$, we parameterize the cycle by time $t$. The winding number $n$ of a limit {cycle} may then be defined as
\begin{equation}
n= \frac{1}{2\pi} \int_{t}^{t+T}dt' \frac{d\psi}{dt'}=\frac{1}{2\pi}[\psi(t+T)-\psi(t)],
\label{eq:winding_number}
\end{equation}
%\begin{eqnarray}
%n=&& \frac{1}{2\pi} \int_{t}^{t+T}dt' \frac{d\psi}{dt'}
%\nonumber \\
%=&&\frac{1}{2\pi}[\psi(t+T)-\psi(t)],
%\label{eq:winding_number}
%\end{eqnarray}
where $T$ is the period and $t$ is an arbitrary point in time. In the case of oscillations, {the} winding number of the limit cycle is $n=0$, since $\psi(t+T)=\psi(t)$. For positive $\Omega > \Omega_C$ (clockwise rotation), rotations are characterized by $n=-1$, since $\psi(t+T)=\psi(t)-2\pi$. Moreover, by comparing Eqs.~(\ref{eq:winding_number}) and (\ref{eq:avg_frq}), we can immediately see that the abrupt drop in cycling frequency \(f\) measured in the original {non-rotating} frame is directly related with the winding number \(n\) {as follows}

\begin{equation}
	f = \frac{\sigma}{2\pi} + \frac{n}{T}.
\end{equation}

In summary, our analysis revealed a topological phase transition (the SPOR transition) in the forced Kuramoto model, from oscillations to wobbly rotations of the complex order parameter \(z\). The transition takes place at $\Omega = \Omega_C$, as the orbit of \(z(\rho,\psi)\) in the rotating frame approaches the singular point \(z=0\), such that \(\mathrm{min}(\rho(t))\to 0\) as shown in Fig.~\ref{fig:figure_4}(a). In the non-rotating frame, the transition is marked by an abrupt drop in the average cycling frequency \(f\), as shown in Fig.~\ref{fig:figure_4}(c).  Numerical analysis of the curvature shows the limit cycles immediately below and above $\Omega_C$ are identical, and related by a simple translation in the order-parameter space, as depicted schematically in Fig.~\ref{fig:figure_4}(d). However, the existence of the singular point \(z=0\) makes this translation impossible, since an adiabatic translation would require the limit cycle to pass through the singular point, which is forbidden (as explained in Sec.~\ref{sec:two}). In addition, the oscillating and rotating phases can be distinguished by a topological characteristic, the winding number \(n\) of the limit cycles, as shown in Fig.~\ref{fig:figure_4}(c). Notably, the  average cycling frequency $f$ measured in the {non-rotating} frame is directly related with the winding number.

\section{Discussion \label{sec:four} }

In this paper we found that, by increasing the frequency or decreasing the strength of an external field, the periodically forced Kuramoto model undergoes an abrupt phase transition from a phase with oscillations to a phase with wobbly rotations of the order parameter. We call this a 'Singular Point Oscillation-to-Rotation' (SPOR) phase transition. In the original {non-rotating} frame, the SPOR transition appears as an abrupt drop in average group frequency of  phase oscillators.

Our analysis of the dynamical behavior of the forced Kuramoto model shows that the SPOR transition can neither be classified as a first nor as a second order phase transition since it does not display critical phenomena characteristic of either transition.  First, according to the classification of phase transitions in Landau's phenomenological theory \cite{landau2013statistical},
%\cite{landau2013statistical,eugene1971introduction}
%In this theory, phase transitions are divided into two broad categories continuous and discontinuous transitions.
second order (continuous) phase transitions are accompanied by symmetry breaking across the transition and a gradual increase of the order parameter in the ordered phase. In the case of the SPOR transition, symmetry breaking is absent. Second, the proximity to the critical point of a continuous transition is signalled by an increase of critical fluctuations that result in critical phenomena, such as a strong increase in susceptibility, correlation length, relaxation rate (known as critical slowing down of dynamics), etc. Our detailed analysis of the dynamical behavior of the forced Kuramoto model near the SPOR transition revealed no anomaly in behavior in the corresponding fixed point, the curvature of the limit cycle, and the relaxation rate. Third, first order  (discontinuous) phase transitions are characterized by an abrupt appearance of order, hysteresis, and a region of metastable states. Moreover, critical phenomena occur close to the critical boundary of metastable states. In our case, hysteresis and metastable states are absent.
Fourth, the absence of critical correlations also suggests the SPOR transition is not a hybrid phase transition, which combines the abrupt appearance of order, as in first order phase transitions, with the absence of hysteresis and the presence of critical fluctuations, as in second order phase transitions (for an example, consider the hybrid transition in the Kuramoto model with frequency-degree correlations \cite{coutinho2013kuramoto}.) Fifth, despite the absence of symmetry breaking, states above and below the critical point of the SPOR transition are topologically distinct states characterized by different winding numbers.

Within the concept of topological phase transition, states of matter can be classified according to their topological properties \cite{mermin1979topological,wen1995topological,hasan2010colloquium,goldman2016topological}.
In contrast to the conventional Landau paradigm of order parameters associated with spontaneous symmetry breaking, topological phases of matter are characterized by non-local topological invariants, such as  Chern numbers for strongly correlated topological phases in ultracold gases \cite{goldman2016topological} {and topological insulators \cite{hasan2010colloquium}}, or winding numbers for topological defects  \cite{mermin1979topological}.
Topological phase transitions occur without symmetry breaking between states with different topological properties.

Our analysis of the SPOR transition {has shown} that a singular point in the  {order-parameter} space,  corresponding to the state with zero order parameter, plays a crucial role in the topological transition. This singular point lies outside the limit cycle of oscillations, but is encircled by limit cycle in the rotating phase.  As a result, the limit cycles corresponding to oscillations and rotation{s} acquire different topological properties and form two topologically distinct classes corresponding to different winding numbers. The limit cycles of oscillations have winding number zero and the limit cycles of rotations have winding number $\pm 1$ depending on the sign of the detuning parameter.
Limit cycles belonging to the same class can be smoothly (adiabatically) deformed one into another, while cycles belonging to different classes {cannot} be deformed smoothly one into another.
Based on this analysis, we conclude that the SPOR transition is a topological phase transition between states with winding numbers 0 and $\pm 1$. Since the transition is topological in origin, it is not caused by a specific bifurcation, and was therefore missed by previous studies based on bifurcation theory \cite{sakaguchi1988cooperative,antonsen2008external,childs2008stability}. {A similar phase transition has also been reported in the Kuramoto model with interacting identical phase oscillators subjected to white noise \cite{zaks2003noise}.}

{As mentioned in the Introduction, topological phases of matter have been found in quantum systems, namely in fractional quantum Hall liquids and topological isolators, see, for example, reviews \cite{wen1995topological,goldman2016topological}. There is an interesting similarity between the SPOR transition in the classical forced Kuramoto model and a transition from the flipper phase to the spinner phase in a mechanical model of topological metamaterials \cite{chen2014nonlinear} (compare our Fig.~\ref{fig:phase_portrait} and Fig.~4(c) in paper \cite{chen2014nonlinear}). Topological states with winding number $\pm 1$ and 0 on Fig.~\ref{fig:phase_portrait} are topologically similar to topological edge states in two-dimensional topolectrical circuits \cite{olekhno2020topological} where bound two photon topological states appear if the topoelectrical circuit has a unit cell characterized by winding number 1, but are absent if the winding number is zero  (see Fig. 3 in paper \cite{olekhno2020topological}). Note that these topological metamaterials and topolelectrical circuits are classical analogs of quantum systems with topological edge states \cite{chen2014nonlinear,olekhno2020topological}.}

%As mentioned above, topological phases of matter have been found in quantum systems, namely in fractional quantum Hall liquids and topological isolators, see, for example, reviews \cite{wen1995topological,goldman2016topological}. The SPOR  transition is an example of a transition between two distinct topological phases in a classical system of interacting phase oscillators with a singular point in the order-parameter space. It is an example of transition from a state where system is moving far from a singular point to a state where the system rotates around this point. There is interesting similarity between the SPOR transition and a  transition from the flipper phase to the spinner phase in a mechanical model of a topological metamaterial \cite{chen2014nonlinear}. This similarity can be seeing, comparing our Fig. \ref{fig:phase_portrait} and Fig. 4(c) in \cite{chen2014nonlinear}.  A similar phase transition has also been reported in the Kuramoto model with interacting identical phase oscillators subjected to white noise \cite{zaks2003noise}, and we believe this kind of topological transition may be found in other classical systems.

{The SPOR  transition is an example of a topological transition between two distinct topological phases with winding numbers $\pm1$ and zero in a classical system of interacting phase oscillators with a singular point in the order-parameter space. We believe this kind of topological transition may be found in other classical systems.} As an illustrative example of a SPOR transition, we may consider a motorboat making circles on the open sea. All circles are topologically equivalent independently of their position. Let us now introduce an observer in the water. The situation becomes crucially different. If the observer is placed outside these circles, then the motorboat oscillates with respect to the observer. The observer can track the motorboat just by turning their neck.  If the motorboat encircles the observer, rotating around them, the observer must also rotate their body to track the motorboat. The critical circle corresponds to the situation when the motorboat passes through the observer. Assuming the person steering the motorboat wishes to avoid committing a crime, passing through the observer is forbidden. In this model, the observer plays the role of the singular point. States with the observer outside and inside the closed tracks created by the motorboat have different topological properties with respect to the observer, {i.e. winding numbers 0 and $\pm 1$, respectively}. A transition from one state to another is an abrupt topological transition.
%Thus, the position of the observer with respect to the motor track impacts his behavior.
%This example is to show that the SPOR transition is a particular case of a general case, when topology becomes important and a change in dynamical behavior is caused by change in topological properties of space of dynamical states with a singular point.

%\end{acknowledgments}
%
%\section*{Data Availability Statement}
%The data that support the findings of this study are available from the corresponding author upon reasonable request.

\section*{Declaration of Competing Interest}
None.

\section*{Credit Author Statement}
All authors contributed equally to this paper. 

\section*{Acknowledgments}
We thank Ricardo Guimarães Dias for useful discussions. This work is funded by national funds (OE), through Portugal's FCT Funda\c{c}\~{a}o para a Ci\^{e}ncia e Tecnologia, I.P., within the scope of the framework contract foreseen in paragraphs 4, 5 and 6 of article 23, of Decree-Law 57/2016, of August 29, and amended by Law 57/2017, of July 19. E. A. P. W. acknowledges the financial support provided by FCT under PhD grant SFRH/BD/121331/2016.

\bibliography{bibliography}

\end{document}